\documentclass[%
reprint,
superscriptaddress,
nofootinbib,
 amsmath,amssymb,
 aps,
 prd,
twocolumn,
]{revtex4-2}

\usepackage[hidelinks,colorlinks,urlcolor=blue,citecolor=blue,linkcolor=blue]{hyperref}
\usepackage[hyphenbreaks]{breakurl}
\usepackage{color,graphicx,graphics,amsfonts,mathrsfs,amsmath,bm,psfrag,natbib,dcolumn,textcase}
\usepackage{enumerate}
\allowdisplaybreaks
\usepackage{epsfig, soul, latexsym, multirow}
\usepackage{comment}
\usepackage{subfigure}
\usepackage{mathtools,nccmath}
\usepackage{bigints}
\usepackage{float}
\usepackage{footmisc}
\usepackage{placeins}
\usepackage[normalem]{ulem}
\usepackage{physics}
\usepackage{setspace}
\usepackage{natbib}
\usepackage{xparse}

\begin{document}

\title{Importance of relativistic pericenter precession in
identifying the presence \\of a third body near eccentric binaries}
\author{Pankaj Saini}
\email{pankaj.saini@nbi.ku.dk}
\affiliation{Niels Bohr International Academy, The Niels Bohr Institute, Blegdamsvej 17, DK-2100, Copenhagen, Denmark}
\affiliation{Center of Gravity, Niels Bohr Institute, Blegdamsvej 17, 2100 Copenhagen, Denmark}
\author{Lorenz Zwick}
\affiliation{Niels Bohr International Academy, The Niels Bohr Institute, Blegdamsvej 17, DK-2100, Copenhagen, Denmark}
\affiliation{Center of Gravity, Niels Bohr Institute, Blegdamsvej 17, 2100 Copenhagen, Denmark}
\author{J\'{a}nos Tak\'{a}tsy}
\affiliation{Niels Bohr International Academy, The Niels Bohr Institute, Blegdamsvej 17, DK-2100, Copenhagen, Denmark}
\affiliation{Center of Gravity, Niels Bohr Institute, Blegdamsvej 17, 2100 Copenhagen, Denmark}
\author{Connar Rowan}
\affiliation{Niels Bohr International Academy, The Niels Bohr Institute, Blegdamsvej 17, DK-2100, Copenhagen, Denmark}
\author{Kai Hendriks}
\affiliation{Niels Bohr International Academy, The Niels Bohr Institute, Blegdamsvej 17, DK-2100, Copenhagen, Denmark}
\affiliation{Center of Gravity, Niels Bohr Institute, Blegdamsvej 17, 2100 Copenhagen, Denmark}
\author{Gaia Fabj}
\affiliation{Niels Bohr International Academy, The Niels Bohr Institute, Blegdamsvej 17, DK-2100, Copenhagen, Denmark}
\affiliation{Center of Gravity, Niels Bohr Institute, Blegdamsvej 17, 2100 Copenhagen, Denmark}
\author{Daniel J. D'Orazio}
\affiliation{Space Telescope Science Institute, 3700 San Martin Drive, Baltimore , MD 21}
\affiliation{Niels Bohr International Academy, The Niels Bohr Institute, Blegdamsvej 17, DK-2100, Copenhagen, Denmark}
\author{Johan Samsing}
\affiliation{Niels Bohr International Academy, The Niels Bohr Institute, Blegdamsvej 17, DK-2100, Copenhagen, Denmark}
\affiliation{Center of Gravity, Niels Bohr Institute, Blegdamsvej 17, 2100 Copenhagen, Denmark}

\date{\today}

\begin{abstract}
Many astrophysical processes can produce gravitational wave (GW) sources with significant orbital eccentricity. These binaries emit bursts of gravitational radiation during each pericenter passage. In isolated systems, the intrinsic timing of these bursts is solely determined by the properties of the binary. The presence of a nearby third body perturbs the system and alters the burst timing. Accurately modeling such perturbations therefore offers a novel approach to detecting the presence of a nearby companion. Existing timing models account for Newtonian dynamics and leading order radiation reaction effects but neglect the higher order post-Newtonian (PN) contributions to the inner binary. In this paper, we present an improved timing model that incorporates conservative PN corrections that lead to the precession of the binary's pericenter. We find that these PN corrections significantly impact the binary's orbital evolution and the timing of the GW burst. In particular, 1PN precession gives rise to distinctive modulation features in the binary's semilatus rectum and eccentricity. These modulations encode valuable information about the presence and properties of the third body, including its mass and distance. Furthermore, unmodeled 1PN effects significantly bias the tertiary's mass and distance. Finally, we assess the detectability of GW bursts from such perturbed systems and demonstrate that the inclusion of PN corrections is crucial for accurately capturing the orbital dynamics of hierarchical triples.

\end{abstract}

\maketitle

\section{Introduction}
Advanced LIGO (LIGO)~\cite{LIGOScientific:2014pky} and Advanced Virgo (Virgo)~\cite{VIRGO:2014yos} have detected gravitational wave (GW) signals from the mergers of more than $200$ compact binary systems during the first four observing runs~\cite{LIGOScientific:2020ibl,KAGRA:2021duu, KAGRA:2021vkt, Nitz:2021zwj,LIGOScientific:2025rsn}. Many formation scenarios have been proposed for these binaries, including isolated binary evolution in the galactic field~\cite{Bavera:2020uch}, dynamical interactions in stellar clusters~\cite{Chatterjee:2016thb, Mandel:2018hfr, DiCarlo:2019pmf}, and binaries in active galactic nuclei (AGN)~\cite{Tagawa:2019osr,Samsing:2020tda,Tagawa:2020jnc, Rowan:2022ehz, Whitehead:2023hmh, Rowan:2025xxb}. Current observations of binary black holes (BBHs) by the LIGO-Virgo-KAGRA (LVK) collaboration suggest that a combination of formation pathways is preferred over a single channel~\cite{LIGOScientific:2020kqk,Bouffanais:2021wcr,PhysRevD.103.023026,Zevin:2020gbd}.

Orbital eccentricity of a binary is a key indicator of its formation scenario. While isolated formation channels typically produce quasi-circular binaries due to efficient circularization through radiation reaction~\cite{PhysRev.136.B1224}, dynamically formed binaries in dense stellar environments -- such as globular clusters and nuclear star clusters -- can retain measurable eccentricity in the LVK frequency band~\cite{Samsing:2013kua, PhysRevD.97.103014, Rasskazov:2019gjw, Zevin:2021rtf,DallAmico:2023neb}. Recent studies have reported weak evidence of eccentricity in a subset of events in GWTC-3 catalog~\cite{Romero-Shaw:2020thy,Gayathri:2020coq,Romero-Shaw:2021ual,OShea:2021faf,Gupte:2024jfe, Morras:2025xfu, Jan:2025fps, Kacanja:2025kpr}. While LIGO is sensitive to BBH eccentricities $\gtrsim 0.05$ (at $10$ Hz GW frequency)~\cite{PhysRevD.98.083028, Favata:2021vhw}, third-generation (3G) detectors -- Cosmic Explorer (CE)~\cite{LIGOScientific:2016wof, Reitze:2019iox} and Einstein Telescope (ET)~\cite{Punturo:2010zz,Sathyaprakash:2011bh} -- with enhanced sensitivity at low frequencies and better overall sensitivity, will be sensitive to $\sim 10$ times smaller eccentricities~\cite{Saini:2023wdk}. 

In this paper, we mainly focus on highly eccentric binaries that emit GW radiation in the form of GW bursts. GW emission from a highly eccentric binary is suppressed for most of the orbit, except during the pericenter passage~\cite{PhysRevD.90.103001,Loutrel:2017fgu,Xuan:2023azh, PhysRevD.85.123005, PhysRevD.110.023020}. A single GW burst can have SNR $\sim 20$ in LIGO~\cite{Loutrel:2020jfx}, whereas in the LISA band, the SNR for a single burst can reach $\mathcal{O}(100)$~\cite{Xuan:2023azh}. The total SNR from a series of bursts roughly scales as $\sqrt{N}$ for $N$ bursts~\cite{Loutrel:2017fgu, Loutrel:2020jfx}. As fast and accurate waveform models remain underdeveloped for highly eccentric binaries, unmodeled burst search methods such as coherent Wave Burst~\cite{Klimenko:2008fu}, power stacking~\cite{Tai:2014bfa}, and Bayeswave~\cite{Cornish:2014kda} could potentially detect these bursts.

Various dynamical astrophysical processes can produce binaries with large eccentricities in the detector frequency band. Galactic nuclei are potential sites for forming highly eccentric binaries. The secular evolution of binaries in the vicinity of a supermassive black hole (SMBH) can lead to the formation of a highly eccentric binaries through binary-single scatterings and Kozai-Lidov oscillations~\cite{2010ApJ...713...90A, 2015ApJ...799..118P,2013MNRAS.431.2155N, Hoang:2017fvh, 2022ApJ...925..178H,Su_2025_kozai, Camilloni:2023xvf}. Around $70\%$ of these BBHs can have eccentricities $\sim 0.1$~\cite{Takatsy:2018euo,  Tagawa:2019osr, 2019MNRAS.488...47F, Tagawa:2020jnc, Samsing:2020tda, Gondan:2020svr}, and a subset of these BBHs can have eccentricity $> 0.9$ in the LVK frequency band~\cite{OLeary:2008myb}. The presence of an AGN disc can further increase the overall merger rates in the galactic center as well as the potential for residual eccentricity \citep{Rowan:2022ehz,Calcino2024, 2025arXiv250803637V, 2025ApJ...982L..13G} and also dephasing from a third object during a binary-single encounter \citep{Rowan:2025xxb} through the BH interactions with the dense gas disc.

If a binary is located in the vicinity of a third body, it affects the GW burst signal in primarily two ways: (1) intrinsic changes in the inner binary's frequency due to tidal perturbations from the third body, (2) Doppler shift due to acceleration of inner binary's center of mass (COM).~\footnote{The Hubble expansion can also cause acceleration of the binary's COM. However, this cosmological effect is expected to be subdominant compared to the peculiar acceleration induced by nearby astrophysical bodies~\cite{PhysRevD.83.044030, PhysRevD.95.044029}.} A notable example is a stellar mass binary orbiting a SMBH, forming a hierarchical triple system. In the Galactic Center, such triple configurations can naturally arise as the stellar mass BBH become bound to the central SMBH~\cite{Grishin:2021hcp}. Tidal perturbations encode information about the intrinsic properties of the triple system, offering a novel method to probe and characterize the presence of a third body. In contrast, Doppler-induced modulations depend on the observer's line of sight and become significant when the inner binary's COM moves appreciably during the observation time. Both of these modulations induce a phase shift in the GW signals relative to that of an isolated binary~\cite{Meiron:2016ipr, Inayoshi:2017hgw, PhysRevD.105.024017, Vijaykumar:2023tjg, Tiwari:2023cpa, Samsing:2024syt, Hendriks:2024zbu,Hendriks:2024gpp,l542-s32g}. Comparing the GW burst timing of a perturbed binary with that of an unperturbed binary thus provides a unique probe of the binary's astrophysical environment in which these binaries form. Therefore, by accurately modeling the influence of a third body on the inner binary, one can detect the presence of a nearby companion and place constraints on their astrophysical environments [e.g.,~\cite{Yunes:2010sm, Laeuger:2023qyz}]. In this work, we mainly focus on the timing modifications induced by tidal effects from a third body.

Recently, Romero-Shaw~$et\; al.$~\cite{Isobel} computed the third body effects on the timing of GW bursts from a highly eccentric binary and investigated the ability of third-generation detectors to constrain the properties of a perturbing massive third body. The timing model presented in ~\cite{Isobel} only accounts for the Newtonian and 2.5PN order effects in the dynamics of the inner binary and neglected the higher order post-Newtonian (PN) effects to simplify their model. Notably, higher order PN effects are not negligible and can significantly impact the GW burst timing and the orbital evolution of the binary. 

More recently, Samsing $et\, al.$~\cite{Samsing:2024syt} performed a set of N-body simulations of triple systems, with leading order conservative (Newtonian) and dissipative (2.5PN) effects. They also incorporated higher order conservative PN terms -- specifically 1PN and 2PN corrections -- that lead to pericenter precession of the inner binary. Their results demonstrate that neglecting these conservative PN effects in inner binary dynamics leads to an overestimation of GW phase shifts between a perturbed and a reference (unperturbed) binary. Furthermore, they found that higher order PN terms introduce additional modulation features in the binary's orbital parameters such as semi-major axis and eccentricity.

In this paper, we bridge the gap between theory and simulation by analytically incorporating conservative PN contributions to the inner binary dynamics, which induce relativistic pericenter precession of the orbit. We derive the 1PN corrections to the GW burst timing and find that these corrections significantly change the timing of the GW burst. Furthermore, 1PN corrections give rise to additional modulation features in the orbital parameters of the inner binary -- features consistent with those observed in the PN simulations~\cite{Samsing:2024syt}. Moreover, in the absence of 1PN precession, the GW burst timing strongly depends on the initial argument of pericenter. Once pericenter precession is accounted for, however, this dependence becomes substantially weaker. These precession-induced effects have important implications for detecting and characterizing third body companions in eccentric binary systems. Therefore, PN corrections are crucial for accurately modeling the GW burst timing and orbital evolution of the binary.

The rest of the paper is organized as follows. In Sec.~\ref{sec:method_of_osculating_orbit}, we describe the method of osculating orbits that is used to solve the perturbed Kepler problem. Section~\ref{sec:parameter space} provides the validity conditions of the timing model based on physical constraints and discusses the allowed region of parameter space for a third body system. In Sec.~\ref{sec:1PN_acceleration}, we derive the evolution equations for the GW burst timing and orbital parameters of inner binary including the effects of pericenter precession. In Sec~\ref{sec:importance of 1PN precession}, we discuss the importance of 1PN corrections in the timing model. Section~\ref{sec:detectability} discusses the detectability of tertiary perturbations by comparing the difference in the timings of perturbed and unperturbed binaries. In Sec.~\ref{sec:conclusion}, we present the conclusions of our study. We use units in which $G=c=1$.

\section{Methods}
While the general three body problem lacks an exact analytical solution, it becomes tractable under the assumption that the third body induces only a weak perturbation on the inner binary. In this perturbative regime, the dynamics can be described using the method of osculating orbital elements.

\subsection{Osculating orbit formalism}\label{sec:method_of_osculating_orbit}
We consider a three body system composed of a highly eccentric inner binary and a perturbing third body on a circular outer orbit. The inner binary consists of bodies $1$ and $2$ with masses $m_1$ and $m_2$. The third body has mass $m_3$ and orbits at a much larger distance $R$ compared to the inner binary separation $r$, such that it induces only a weak perturbation to the inner binary's motion. We assume that the inner and outer orbits are coplanar. 

\begin{figure}
    \centering
{\includegraphics[width=0.48\textwidth]{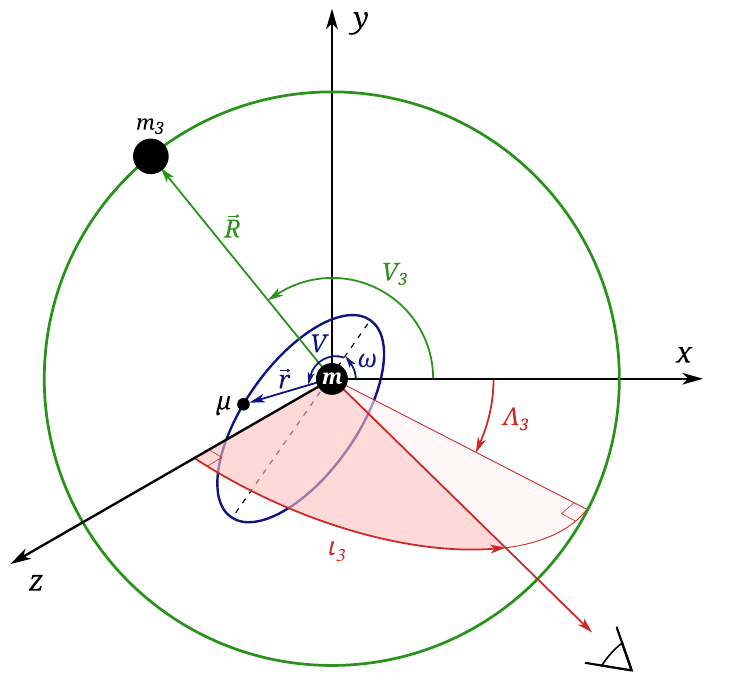}}
 \caption{Geometric representation of a triple system. The inner binary follows an eccentric orbit with reduced mass $\mu$ orbiting the total mass $m$ at a separation $r$. The third body of mass $m_3$ is located at a much larger distance $R \gg r$ and moves on a circular outer orbit. The angle $V$ denotes the true anomaly of the inner orbit, and $\omega$ is the longitude of pericenter. The true anomaly of the outer orbit is $V_3$. Both orbits are assumed to be coplanar. The inclination angle $\iota_3$ describes the orientation of the system relative to the detector.}
\label{fig:schematic}
\end{figure}

Figure~\ref{fig:schematic} shows the geometric illustration of a triple system. A test particle of reduced mass $\mu = m_1m_2/(m_1+m_2)$ moves in the gravitational potential of a total mass $m=m_1+m_2$, located at the center of mass of the inner binary. We denote the true anomaly of the inner orbit by $V=\phi-\omega$, where $\phi$ is the orbital phase and $\omega$ is the angle of the pericenter. The true anomaly of the outer orbit is represented by angle $V_3$. The orbital eccentricity is denoted by $e$, the semi-major axis by $a$, and $p=a(1-e^2)$ represents the semi-latus rectum of the inner binary. The orientation of the source with respect to the detector is represented by inclination angle $\iota_3$. We assume $\iota_3 = 0$ throughout, except in Sec.~\ref{sec:COM_motion}.

The system is an example of a \textit{perturbed Kepler problem}. To solve it, we employ the \textit{method of osculating orbits}~\cite{Poisson_Will_2014} \footnote{See also~\cite{2025arXiv250806245G} for a unified perturbed Kepler framework incorporating a large class of deviations from Keplerian binary due to environmental effects.}. Due to the presence of a small perturbing force, the otherwise constant orbital elements of the binary evolve slowly over time. As a result, the dynamics of the binary can be approximated by a series of instantaneous Keplerian orbits whose parameters evolve secularly due to the perturbing force. The relative acceleration between two bodies $(\bm{a} = \bm{a}_1-\bm{a}_2)$ can be written in the form 
\begin{align}
      \bm{a} = -\frac{m}{r^2}\bm{n} + \bm{f} \,,  
\end{align}
where $\bm{n} =  \bm{r}/r$ is a unit vector that points from body 2 to body 1, $|\bm{r}| = r$ is the separation between bodies, $\bm{f}$ is a perturbing force per unit mass, and the radial distance to the test particle is given by
\begin{equation}
        r = \frac{p}{1+e \cos V} \,.
\end{equation}
The perturbing force can be decomposed into a vectorial basis for the orbital plane
\begin{equation}
    \bm{f} = \mathcal{R} \bm{n} + \mathcal{S} \bm{\lambda} + \mathcal{W} \bm{e_z} \,.
\end{equation}
The components of the perturbing force in ($\bm{n}, \bm{\lambda}, \bm{e_z}$) basis are
\begin{equation}\label{eq:components_of_perturbing_force}
    \mathcal{R} = \bm{f}\cdot\bm{n}, \;\; \mathcal{S} = \bm{f} \cdot \bm{\lambda}, \;\; \mathcal{W} = \bm{f}\cdot\bm{e_z} \,,
\end{equation}
The following equations describe the vectorial basis: 
\begin{subequations}
\begin{align}
\bm{n} & = [\cos(V+\omega), \sin(V+\omega), 0] \,, \\
\bm{\lambda} & = [-\sin(V+\omega), \cos(V+\omega), 0] \,, \\
\bm{e_z} &= [0, 0, 1] \,.
\end{align}
\end{subequations}
We assume that the tertiary is located in the plane of the inner binary and $V_3$ is the phase of the outer orbit. The coordinates of the third body are given by the following vector  
\begin{equation}
    \bm{N} = [\cos V_3, \sin V_3, 0] \,.
\end{equation}
where $\bm{N} = \bm{R}/R$. The orbital evolution is governed by the following system of osculating equations~\cite{Poisson_Will_2014}:
\begin{subequations}\label{eq:osculating_equations}
    \begin{align}
\frac{dp}{dV} & \simeq 2 \frac{p^3}{m} \frac{1}{(1+e \cos V)^3} \mathcal{S} \,, 
 \\
  \frac{de}{dV} & \simeq \frac{p^2}{m}\bigg[\frac{\sin V}{(1+ e \cos V)^2 }\mathcal{R} 
  + \frac{2 \cos V + e (1+ \cos^2 V)}{(1+ e \cos V)^3} \mathcal{S} \bigg] \,, 
\\
 \frac{d\omega}{dV} & \simeq \frac{1}{e} \frac{p^2}{m} \bigg[-\frac{\cos V}{(1+ e \cos V)^2} \mathcal{R} 
 + \frac{2 + e \cos V}{(1+ e \cos V)^3}\sin V \mathcal{S} \bigg] \,,
\\
 \frac{dt}{dV} & \nonumber \simeq \left(\frac{p^3}{m} \right)^{1/2} \frac{1}{(1+e \cos V)^2} \times \Bigg\{1- \frac{1}{e}\frac{p^2}{m} 
 \\
 &\Bigg[\frac{\cos V}{(1+e \cos V)^2} \mathcal{R} - \frac{2+e\cos V}{(1+ e \cos V)^3} \sin V \mathcal{S}\Bigg] \Bigg\} \,.
   \end{align}
\end{subequations}
These are first-order differential equations and require initial values $\{p_0, e_0, \omega_0, t_0\}$ for given $\{m,\mu, m_3, R\}$.

\subsection{Constraints on the parameter space}\label{sec:parameter space}
The applicability of the timing model is subject to certain physical constraints. In the following, we outline the key conditions that define the valid regime of our model.

\subsubsection{Stability}
For the coplanar and circular outer orbit, the stability criterion for a triple system is given by~\cite{2001MNRAS.321..398M,Blaes:2002cs}
\begin{equation}\label{eq:stability}
    \frac{R}{m_3} \gtrsim 2.8 \frac{(p/m)}{(m_3/m)^{3/5} (1-e^2)} \,.
\end{equation}
This criterion ensures long-term dynamical stability by preventing the third body from disrupting the inner binary. Note that mutually inclined orbits are expected to be more stable~\cite{Blaes:2002cs}, so Eq.~\eqref{eq:stability} provides a more conservative stability limit. [See also~\cite{2022MNRAS.516.4146V, 2022PASA...39...62T} for outer eccentric orbits and various mass ratios.]

\subsubsection{Chirping outer binary}
We neglect the PN corrections and radiation reaction effects on the outer orbit. This assumption is justified because we assume that the third body is located at a distance much larger than the binary separation, i.e., $R \gg r$. To leading order, the radius of a circular outer orbit shrinks according to the following expression~\cite{PhysRev.136.B1224}:
\begin{equation}
    R(t) = \left[R_0^4 -\frac{256}{5} \eta_{\rm out} (m+m_3)^3 t\right]^{1/4}\,,
\end{equation}
where $\eta_{\rm out} = (m_3 m)/(m + m_3)^2$. To quantify the effect of radiation reaction on the outer orbit, we require that the radius of the outer orbit does not evolve significantly during the observation time, i.e., $\delta R/R_0 \lesssim 1\%$~\cite{Chandramouli:2021kts}.
\begin{equation}
\frac{R}{m_3} \gtrsim \left(\frac{m}{m_3}\right)  \left[1280 \;\frac{\eta_{\rm out}}{m} \left(1+ \frac{m_3}{m}\right)^3 P_{\rm obs} \right]^{1/4} ,
\end{equation}
where $P_{\rm obs}$ is the observation time. 
\subsubsection{Orbital timescales}

Since the third body is located at a much larger distance, the orbital timescales of the two orbits differ significantly. Specifically, the inner binary completes its orbit much more rapidly than the outer companion, such that $T_{\rm orb, in} \ll T_{\rm orb, out}$. This leads to the following constraint:
\begin{subequations}
\begin{align}
    2\pi \left[\frac{p^3}{m (1-e^2)^3}  \right]^{1/2} \ll  2 \pi \left[\frac{R^3}{(m_3 + m)}\right]^{1/2} \,,
\end{align}
\begin{align}
    \frac{R}{m_3} \gtrsim \frac{(p/m) \left[1+(m_3/m)\right]^{1/3}}{(m_3/m)(1-e^2)} \,.
\end{align}
\end{subequations}
Together, these constraints define the regime of validity for our timing model. 

\begin{figure}
    \centering
    \includegraphics[width=0.48\textwidth]{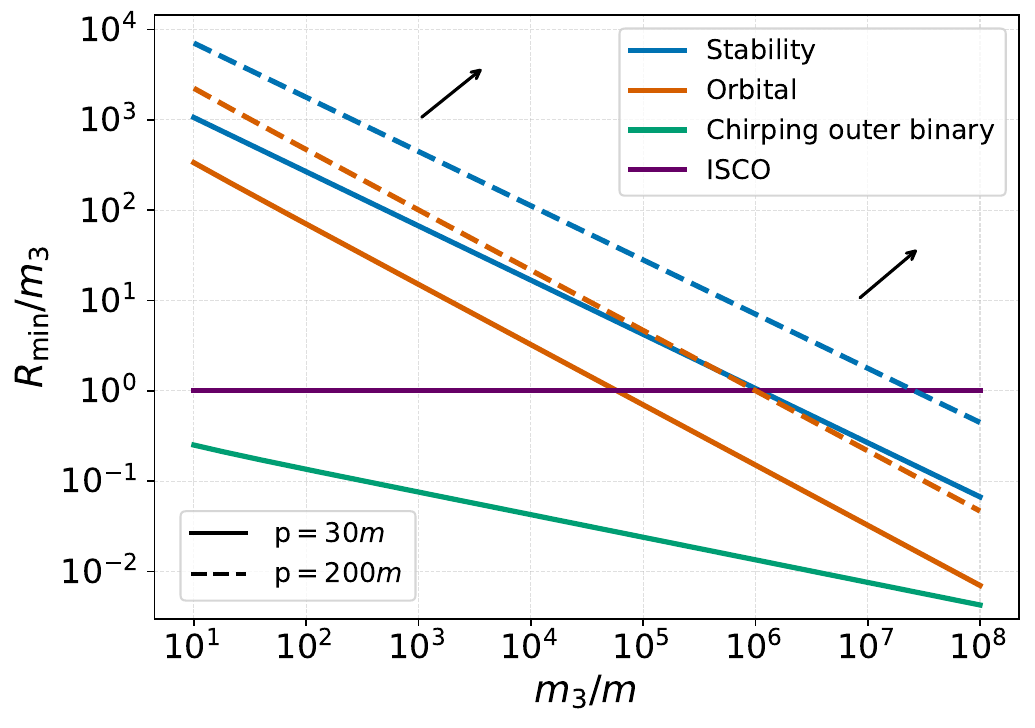}
   \caption{Allowed values of tertiary distance $R$ and mass $m_3$ for various physical constraints discussed in Sec.~\ref{sec:parameter space}. Different colors represent different physical conditions. Solid lines denote $p=30 m$ and dashed lines show $p=200m$. We choose an equal mass binary with $m=30M_{\odot}$ and $e =0.99$. The horizontal line represents the radius of the innermost stable circular orbit for a maximally spinning black hole $(R_{\rm min} = m_3)$. The arrows denote the region of allowed parameters.}
   \label{fig:validity_condition}
   \end{figure}

Figure~\ref{fig:validity_condition} shows these physical constraints in different colors as a function of $m_3$ and $R$. We choose an equal mass binary with $m=30 M_{\odot}$ , $e=0.99$. The solid and dashed lines represent $p = 30 m$ and $p=200 m$, respectively. The horizontal line represents the radius of the innermost stable circular orbit (ISCO) $R_{\rm min}=m_3$ for a maximally spinning black hole. The region above these lines is the allowed parameter space. Note that as the value of $p$ increases, the third body of mass $m_3$ should be located at a larger $R$ otherwise the third body effects can lead to instability in the system. 

In eccentric systems, GW emission occurs not only at the fundamental orbital frequency but also at its integer harmonics. For highly eccentric binaries, the radiation is dominated by higher harmonics, leading to a spectral shift toward higher GW frequencies~\cite{PhysRev.131.435}. The peak GW frequency, corresponding to the harmonic with the maximum power, can be estimated using the following expression~\footnote{Note that this expression becomes inaccurate for low eccentricities. For low eccentricity binaries, Ref.~\cite{Hamers:2021eir} provided an improved numerical fit for $f_{\rm peak, GW}$. Since we consider only highly eccentric binaries in this study, Eq.~\eqref{eq:peak_freq} is good enough for our purpose.}~\cite{Wen:2002km}: 
\begin{equation}\label{eq:peak_freq}
    f_{\rm peak, GW} = n_{\rm peak}(e) f_{\rm orb} \equiv (1+e)^{1.1954} \frac{m^{1/2}}{\pi p^{3/2}} \,.
\end{equation}
Here $n_{\rm peak}$ is the peak harmonic number and $f_{\rm orb}$ is the orbital frequency. Most of the sources considered in this work emit GWs with peak frequencies that lie within the sensitivity band of ground-based detectors such as LIGO, CE, and ET. For example, an equal mass binary with $m=30 M_{\odot}$, $p = 200m$, $e_0=0.99$ will have a peak frequency of $\sim 2$Hz, lower cut-off frequency of Einstein Telescope. 

\subsection{Timing model with 1PN corrections}\label{sec:1PN_acceleration}
Our goal is to determine the GW burst timing under the influence of a third body, including the 1PN corrections to the inner binary dynamics. Since 1PN corrections represent the leading order effects that lead to the precession of the pericenter, here we focus on the 1PN corrections. However, for completeness, we provide the 2PN corrections to the timing model in the Appendix~\ref{app:2PN corrections}. 

Recall that we consider a hierarchical three body system composed of a highly eccentric inner binary and a nearby third body on a circular outer orbit. The equations of motion of the inner binary can be expressed as follows:
\begin{align}\label{eq:total_acceleration}
    \bm{a} &= -\frac{m}{r^2}\bm{n} + \bm{a}_{\rm 1PN} - \frac{m_{3} r}{R^3} \bigg[\bm{n} - 3 (\bm{n} \cdot\bm{N}) \bm{N} + \mathcal{O}(r/R) \bigg] \,,
\end{align}
where the first term is the Newtonian acceleration, the second term is the 1PN acceleration, and the third term is the leading order tidal perturbation due to the third body.\footnote{In Eq.~\eqref{eq:total_acceleration}, we neglect the cross term between 1PN acceleration and the third body. This term vanishes for coplanar orbits~\cite{Will:2013cza}, as we have considered in our study.} The 1PN acceleration is given by~\cite{Poisson_Will_2014, PhysRevD.42.1123}
 \begin{align}
   \bm{a}_{\rm 1PN} & = -\frac{m}{r^2} \nonumber \bigg\{\bigg[(1+3 \eta)v^2 - \frac{3}{2}\eta \Dot{r}^2-2(2+\eta)\frac{m}{r}\bigg]\bm{n} \\ 
  & - 2(2-\eta)\Dot{r}\bm{v} \bigg\}\,,
 \end{align}
where $\eta = m_1 m_2/m^2$ is the binary symmetric mass ratio. The third body effects and the 1PN acceleration in Eq.~\eqref{eq:total_acceleration} constitute the perturbing force. 

The velocity vector $\bm{v}$ is defined as 
\begin{equation}
    \bm{v} = \Dot{r} \bm{n} + r \Dot{\phi} \bm{\lambda} 
\end{equation}
with  
\begin{subequations}
    \begin{align}
        \Dot{r} &= \left(\frac{m}{p}\right)^{1/2} e \sin V \,,
  \\
         \Dot{\phi}& = \left(\frac{m}{p^3} \right)^{1/2} (1+e \cos V)^2 \,,
    \end{align}
\end{subequations}
and $|\bm{v}| = v$. The components of the perturbing force can be calculated using Eq.~\eqref{eq:components_of_perturbing_force} and are given by the following equations.
\begin{subequations}
   \begin{align}
    \mathcal{R} & = \frac{m^2}{4 p^3}\nonumber \left(1+ e \cos V\right)^2 \Big[e^2 \left(4-13 \eta \right) + 4 \left(3-\eta \right) \,  
    \\
       & + 8 e \left(1-2 \eta \right) \cos V - e^2 \left(8-\eta \right) \cos (2V)\Big] \, \nonumber 
       \\ 
     & + \frac{m_3 p}{2 R^3} \frac{\left[1+3 \cos(2(V-V_3+\omega))\right]}{1+e \cos V} \,, 
     \\
     \mathcal{S} &= 2 \frac{e m^2 (2 - \eta)}{p^3} (1+e \cos V)^3 \sin V \, \nonumber
     \\
    &-\frac{3 m_3 p}{2 R^3}  \frac{\sin(2 (V-V_3+\omega))}{1+e \cos V} \,, 
    \\
 \mathcal{W} &= 0 \,.
  \end{align}
\end{subequations}
For a highly eccentric orbit, gravitational radiation is primarily emitted during the pericenter passage, while the emission is significantly suppressed during the remainder of the orbit. Assuming that the orbital elements remain constant throughout the orbit and vary only during the pericenter passage, the secular change in orbital elements $\bm{x} = [p,e,\omega]$ over a complete orbit is given by~\cite{Poisson_Will_2014}
\begin{align}
    \bm{x}_i - \bm{x}_{i-1} = \int^{P}_0 dt \left(\frac{d\bm{x}}{dt}\right)_{\bm{x}=\bm{x}_{i-1}} = \int^{2\pi}_0 dV\left(\frac{d\bm{x}}{dV}\right)_{\bm{x}=\bm{x}_{i-1}} .
\end{align}
The timing between consecutive bursts can be calculated following a similar procedure. The leading order radiation reaction effects (2.5PN) are incorporated in the timing model by following Ref.~\cite{Arredondo:2021rdt}. The complete timing model including the effects of a perturbing third body, the effects of 1PN precession, and 2.5PN radiation reaction effects is given by the following system of equations:
\begin{widetext}
\begin{align}
\label{eq:p}
    p_i &= p_{i-1}\bigg[1-\frac{128 \pi}{5}\eta\bigg(\frac{m}{p_{i-1}} \bigg)^{5/2} \bigg(1+\frac{7}{8} e_{i-1}^2 \bigg) + 15 \pi \left(\frac{m_3}{m} \right) \bigg(\frac{m}{R}\bigg)^3 \bigg(\frac{m}{p_{i-1}} \bigg)^{-3} \frac{e_{i-1}^2 \sin[2(V_{3} -\omega_{i-1})]}{(1-e_{i-1}^2)^{7/2}}\bigg] \,,
\\ \label{eq:e}
    e_i &= e_{i-1}\bigg[1-\frac{608 \pi}{15}\eta\bigg(\frac{m}{p_{i-1}} \bigg)^{5/2} \bigg(1+\frac{121}{304} e_{i-1}^2 \bigg) - \frac{15 \pi}{2} \left(\frac{m_3}{m} \right) 
  \bigg(\frac{m}{R}\bigg)^3 \bigg(\frac{m}{p_{i-1}} \bigg)^{-3} \frac{\sin[2(V_{3} -\omega_{i-1})]}{(1-e_{i-1}^2)^{5/2}}\bigg] \,,
\\ \label{eq:omega}
    \omega_i &= \omega_{i-1} + \frac{6 \pi m}{p_{i-1}} +\frac{3 \pi}{2} \left(\frac{m_3}{m}\right) \bigg(\frac{m}{R} \bigg)^3 \bigg(\frac{m}{p_{i-1}} \bigg)^{-3} \frac{1+5\cos[2(V_3-\omega_{i-1})]}{(1-e_{i-1}^2)^{5/2}} \,,
\\ \label{eq:time}
    t_i &=  \nonumber t_{i-1} + \frac{2\pi}{m^{1/2}} \left[\frac{p_{i-1} + \eta \frac{m^{5/2}}{p_{i-1}^{3/2}} A(e_{i-1}, p_{i-1}, \eta, m)}{1-e_{i-1}^2 + \eta \big(\frac{m}{p_{i-1}} \big)^{5/2} B(e_{i-1})} \right]^{3/2}
    \\
    & \times \nonumber \Biggl[1+ \frac{(1-e^2_{i-1}) m \left[(20-10 \eta)(1- \sqrt{1-e^2_{i-1}}) + e_{i-1}^2 (2+3\eta)\right]}{2 e_{i-1}^2 p_{i-1}}
    \\
   & +\left(\frac{m_3}{m}\right) \left(\frac{m}{R}\right)^3 \left(\frac{m}{p_{i-1}} \right)^{-3} \frac{5(4+3 e_{i-1}^2) + (96+ 51 e_{i-1}^2) \cos[2(V_3-\omega_{i-1})]}{16(1-e_{i-1}^2)^3} \Biggr] \,.
\end{align}
\end{widetext}
The functions $A(e,p, \eta, m)$ and $B(e)$ in Eq.~\eqref{eq:time} can be found in Eqs.~(63)-(66) of Ref.~\cite{Arredondo:2021rdt}. Note that the tidal correction to these equations scales as $(m/p)^{-3} \propto v^{-6}$, that corresponds to a -3PN correction. Also note that Eqs.~\eqref{eq:p} and ~\eqref{eq:e} only contain 2.5PN effects and third body effects, as there are no 1PN corrections to $p$ and $e$. Equation~\eqref{eq:omega} includes 1PN effects and third body effects, as there are no radiation-reaction corrections to $\omega$. It is important to emphasize that these equations consists of a system of coupled equations, meaning that modifications to any individual equation will affect the others.

\section{Importance of the 1PN precession}\label{sec:importance of 1PN precession}

\subsection{Impact of 1PN corrections on GW burst timing}
\begin{figure}
    \centering
{\includegraphics[width=0.48\textwidth]{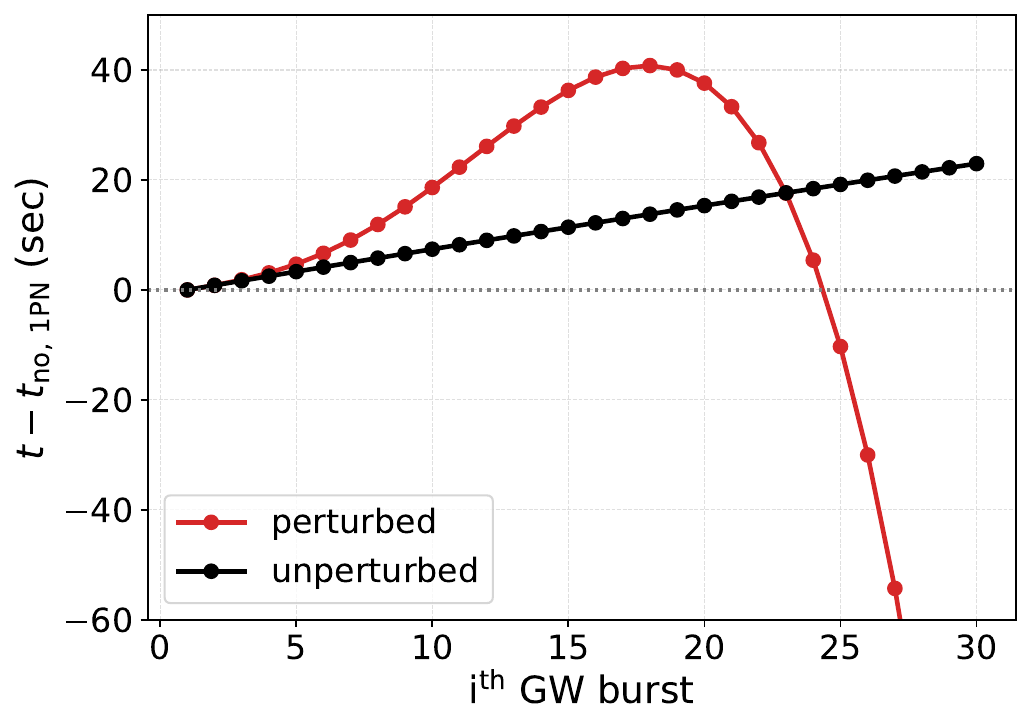}}
 \caption{Difference in GW burst timings with 1PN ($t$) and without 1PN ($t_{\rm no, 1PN}$) corrections as a function of number of bursts. Each dot represents the difference in timing recorded at each pericenter passage. The black and red dots denote the unperturbed and perturbed binary, respectively. The horizontal dotted line shows the typical detection threshold ($0.1$ sec) for the Einstein Telescope. The initial binary parameters are $m=30 M_{\odot}, e_0=0.99, p_0=200 m, \omega_0=0, V_{3,0} =\pi/3$ perturbed by a third body of mass $m_3=10^6 m$ located at $R=20 m_3$. For both the unperturbed binary and perturbed binary, the 1PN corrections significantly impact the GW burst timings.}
\label{fig:timing}
\end{figure}  

In Sec.~\ref{sec:parameter space}, we discussed the parameter constraints of our timing model. Here, we consider a triple system within the validity region of our model and study the evolution of the burst timing and the binary parameters. We choose an inner binary with $m = 30M_{\odot}, e_0 = 0.99, p_0 = 200 m, \omega_0 = 0, V_{3,0} = \pi/3$ perturbed by a third body of mass $m_3 = 10^6 m \; (3 \times 10^7 M_{\odot})$ located at $R = 20 m_3$.

To quantify the effect of 1PN corrections on GW burst timing, we compare the burst times obtained with 1PN corrections to those computed without them. Figure~\ref{fig:timing} shows the difference in timings with 1PN corrections and without 1PN corrections as a function of the number of bursts. Each dot represents the difference in timing during each pericenter passage. The black dots represent the unperturbed binary, and the red dots denote the perturbed binary. For the unperturbed binary, the difference monotonically increases and becomes as large as $\sim 20$ sec at $30^{\rm th}$ burst, which is much greater than the measurement uncertainty in the GW burst timing of 3G detectors $(\sim 0.1\; \text{sec})$~\cite{Isobel}. The perturbed burst timing is initially increasingly delayed compared to the unperturbed case, before reversing and the timing difference becomes significantly negative. The difference becomes as large as $\sim 160$ sec at the end of the $30^{\rm th}$ sec (not shown in the plot). This is because $t_{\rm no, 1PN}$ starts to increase at a faster rate compared to $t$. As the number of bursts increases further, the mismatch between two timings continues to grow in opposite directions.

\begin{figure*}
    \centering
    \begin{subfigure}{\includegraphics[width=0.485\textwidth]{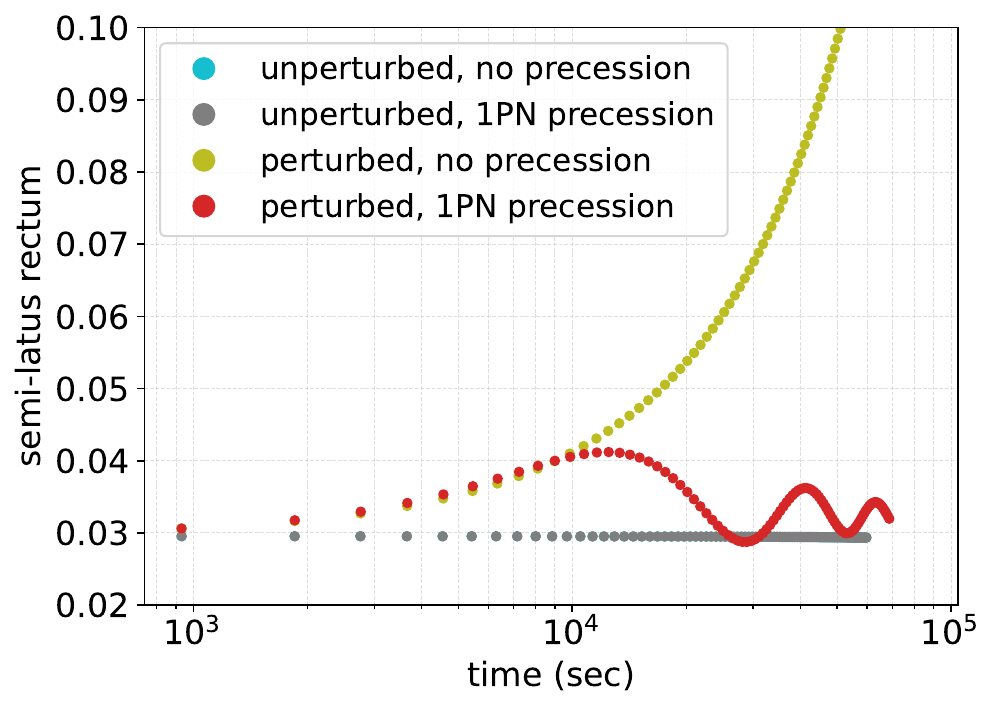}}
    \end{subfigure}
   \begin{subfigure}{\includegraphics[width=0.485\textwidth]{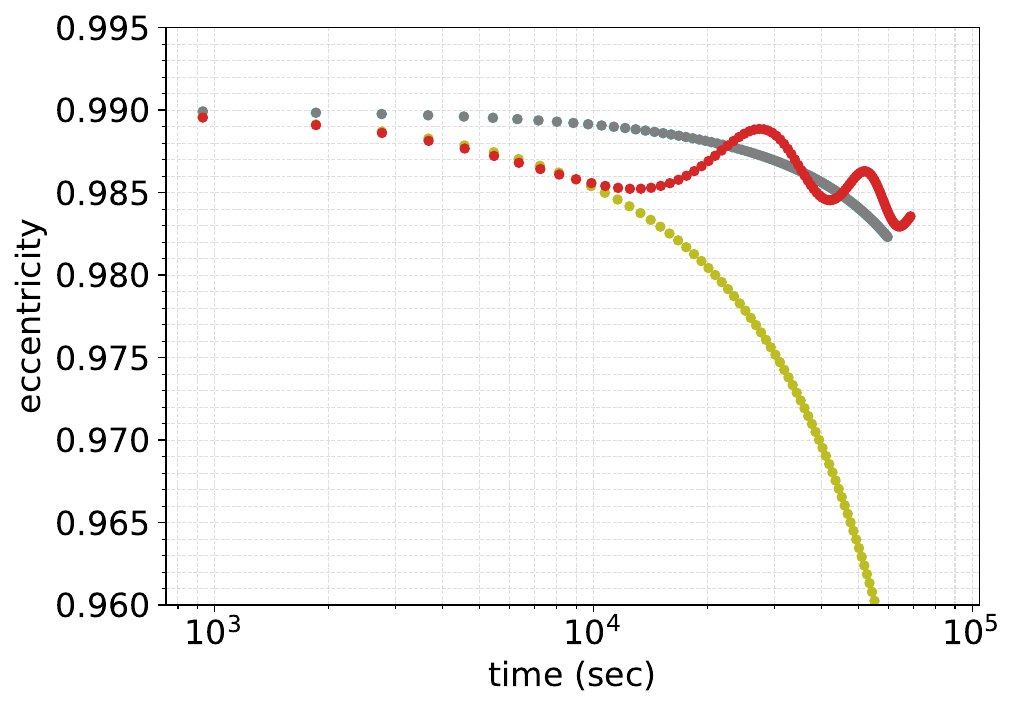}}
   \end{subfigure}
   \begin{subfigure}{\includegraphics[width=0.47\textwidth]{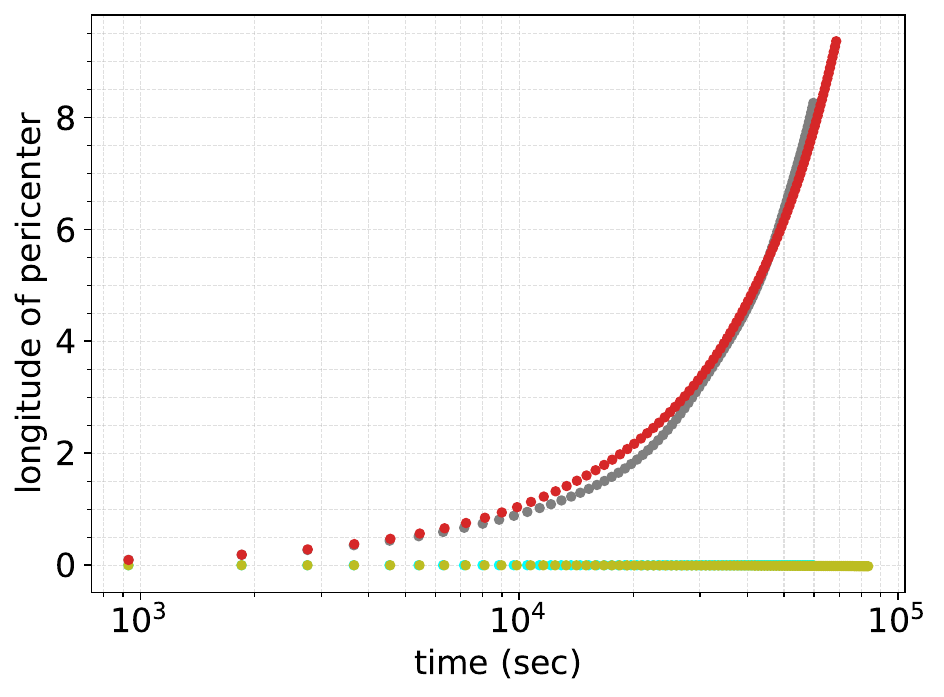}}
   \end{subfigure}
   \caption{Evolution of the semi-latus rectum, eccentricity, and longitude of pericenter of the inner binary. The properties of the system are the same as those in Fig.~\ref{fig:timing}. Different colors represent different physical scenarios. The perturbed binary with 1PN precession shows distinctive modulation features in $p$ and $e$. The longitude of the pericenter increases rapidly with time when 1PN precession is included.}
   \label{fig:p_and_e_modulations_with_time}
   \end{figure*}

In general, the pronounced difference between the perturbed and unperturbed cases arises from the interplay between perturbative corrections due to the third body and the 1PN acceleration terms for the inner binary. Over a pericenter precession timescale, 1PN effects can accumulate and lead to large shifts in the argument of the pericenter, eventually reaching values that are formally non-linear. This transition occurs when $\dot{\omega}_{\rm 1PN} \times t_{\rm obs} \approx 1$ ($\dot{\omega}_{\rm 1PN}$ is the 1PN correction to $\dot{\omega}$ and $t_{\rm obs}$ is the total observation time), corresponding to a significant fraction of a full precession cycle. In this regime, 1PN corrections can exert a disproportionate influence on the GW burst timing, as discussed in \cite{Will:2013cza}. Figure~\ref{fig:timing} clearly shows this transition: at early times (e.g., before the $ \sim 10^{\rm th}$ burst), the system remains in the linear regime, where the 1PN-induced shifts in $\omega$ are relatively small. Beyond this point, the evolution enters the non-linear regime, where cumulative precession effects substantially impact the timing behavior. Therefore, the 1PN corrections significantly impact the burst timing and are crucial for accurately predicting the timings of both perturbed and unperturbed binaries. 

\subsection{Modulations in the binary parameters}\label{sec:modulations}
In Figure~\ref{fig:p_and_e_modulations_with_time}, we plot the binary's semilatus rectum, eccentricity, and longitude of the pericenter as a function of time. We choose the same system parameters as those in Fig.~\ref{fig:timing}. Different colors represent different physical scenarios. For an unperturbed binary, the evolution of $p$ and $e$ is almost identical. The cyan and the gray dots are on top of each other. With 1PN precession, the longitude of the pericenter increases significantly with time for both the perturbed and unperturbed binary. For the perturbed binary, the evolution of $p$ and $e$ is significantly different with and without 1PN precession. In the absence of 1PN precession, $p$ and $e$ do not show any oscillations. However, when 1PN precession is included, it gives rise to periodic oscillations in $p$ and $e$. Notice that, in the absence of precession, $p$ increases and $e$ decreases rapidly. When precession is included, both $p$ and $e$ vary within a small range. It is important to emphasize that the modulations are only present when the binary is perturbed and 1PN precession is included. Therefore, these modulations in the orbital parameters represent the unique signatures for the presence of a third body near the binary. It is important to note that the magnitude and shape of these modulations depend on the properties of the triple system. 

\begin{figure*}
    \centering 
    \subfigure[] 
    { 
    \label{fig:errors_m3R3_m3R3}
    \includegraphics[width=0.48\textwidth]{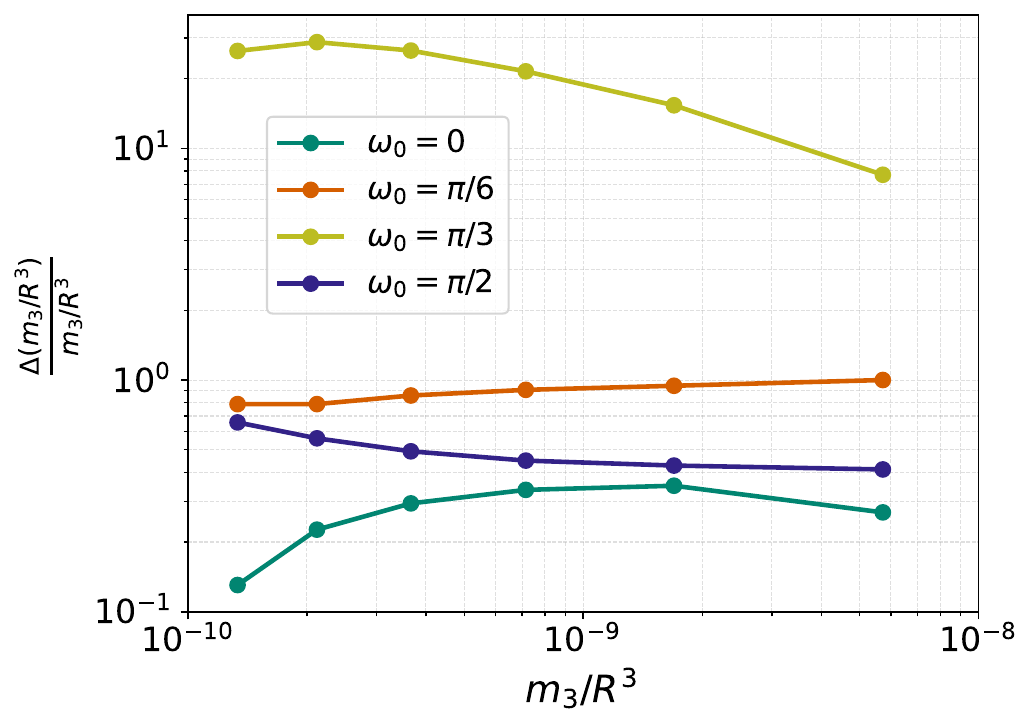}
    }
   \subfigure[]
 { \label{fig:errors_m3R3_e0}
\includegraphics[width=0.48\textwidth]{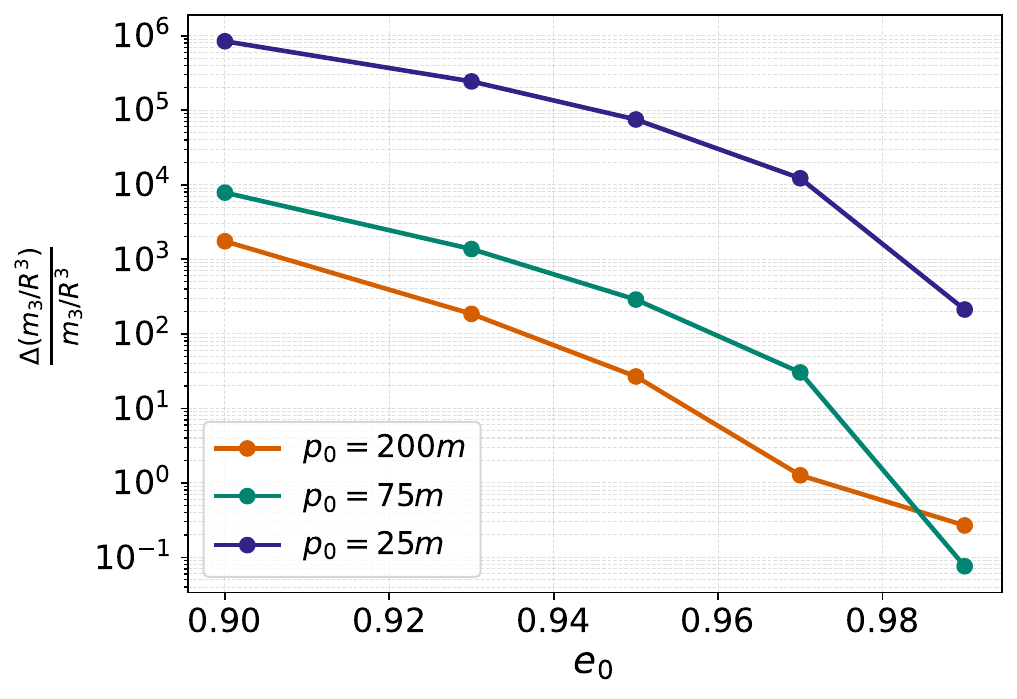}}
\caption{{\it Left:} fractional errors on $m_3/R^3$ due to neglecting 1PN corrections as a function of $m_3/R^3$ and for different values of $\omega_0$. Other binary parameters are the same as those in Fig.~\ref{fig:timing}. {\it Right:} errors as a function of $e_0$ for different values of $p_0$ and $\omega_0 = 0$. The unmodeled 1PN corrections in the timing model lead to significantly large errors in the tertiary mass and distance.}
\label{fig:errors}
\end{figure*}  

\subsection{Bias in the tertiary's mass and distance due to neglecting 1PN effects}
We quantify the systematic errors in the mass and distance of the third body due to neglecting 1PN corrections in the timing model. We do so in a simplified way by fitting the model without 1PN corrections to the model with 1PN corrections. In particular, we minimize the square of the difference between the two models summed over the first $40$ GW bursts and vary the combination of $m_3/R^3$. Since $m_3$ and $R$ are highly degenerate parameters, we obtain errors in $m_3/R^3$. The errors highlight a region of parameter space in which the 1PN corrections are most important. 

Figure~\ref{fig:errors_m3R3_m3R3} shows the fractional systematic errors $\Delta (m_3/R^3)/ (m_3/R^3)$, where $\Delta (m_3/R^3)$ is the difference between the true and best-fit value of $m_3/R^3$. Different colors represent various initial values of $\omega_0$. We fix the initial values of the binary parameter to $p_0=200m$, $e_0=0.99$ and $V_{3,0}=\pi/3$. The errors strongly depend on the initial values of $\omega_0$ and can be significantly large, varying from $\mathcal{O}(10^{-1})$ to $\mathcal{O}(10)$. Note that when $\omega_0=V_{3,0}=\pi/3$, the errors are $\sim 10$ times larger. This is because when there is no precession, the binary's pericenter is always aligned with the third object at each orbit, overestimating the influence of the third object. When 1PN precession is included, the pericenter keeps shifting away from the third object, reducing the influence of the third body. Therefore, the errors are largest when $\omega_0=V_{3,0}$.  

In Fig.~\ref{fig:errors_m3R3_e0}, we plot the errors in $m_3/R^3$ as a function of $e_0$ for different values of $p_0$. We fix $\omega_0=0$ in this plot. The errors decrease as the value of $e_0$ increases. This is because, for fixed values of $p_0$, increasing the value of $e_0$ makes the binary wider for which the tidal effects of the third body dominate over the 1PN precession. Therefore, errors due to precession decrease as $e_0$ increases.  Further, when the value of $p_0$ decreases for a fixed value of $e_0$, the binary becomes more relativistic and 1PN corrections become more important, and therefore, errors increase significantly. For $e_0 = 0.99$, the errors are $\sim 0.1 \mbox{--} 10^2$. These errors are significantly larger for $e_0 = 0.9$ and range from $\sim 10^3$ to $10^6$. In conclusion, neglecting 1PN precession leads to significant errors in the mass and distance of the third body. Therefore, it is crucial to incorporate the 1PN corrections in the timing model.

\section{Detectability of the perturbing third body}\label{sec:detectability}
In this section, we discuss the detectability of the perturbing object. To quantify whether a binary is detectable, we choose the same detection criterion as in~\cite{Isobel}; i.e., we compare the difference in the perturbed and unperturbed burst times $(\Delta t)$ with the approximate measurement uncertainty on the burst time $(\sigma_{\rm tb})$. If $\Delta t > \sigma_{\rm tb}$, the tertiary effects are detectable. We choose $\sigma_{\rm tb} \approx 0.1$ sec which is the mean average standard deviation in burst time posterior samples for the Einstein Telescope~\cite{Isobel}. Note that $\sigma_{\rm tb}$ depends on the properties of the triple system and the given detector. 

\begin{figure}
    \centering
    \includegraphics[width=0.48\textwidth]{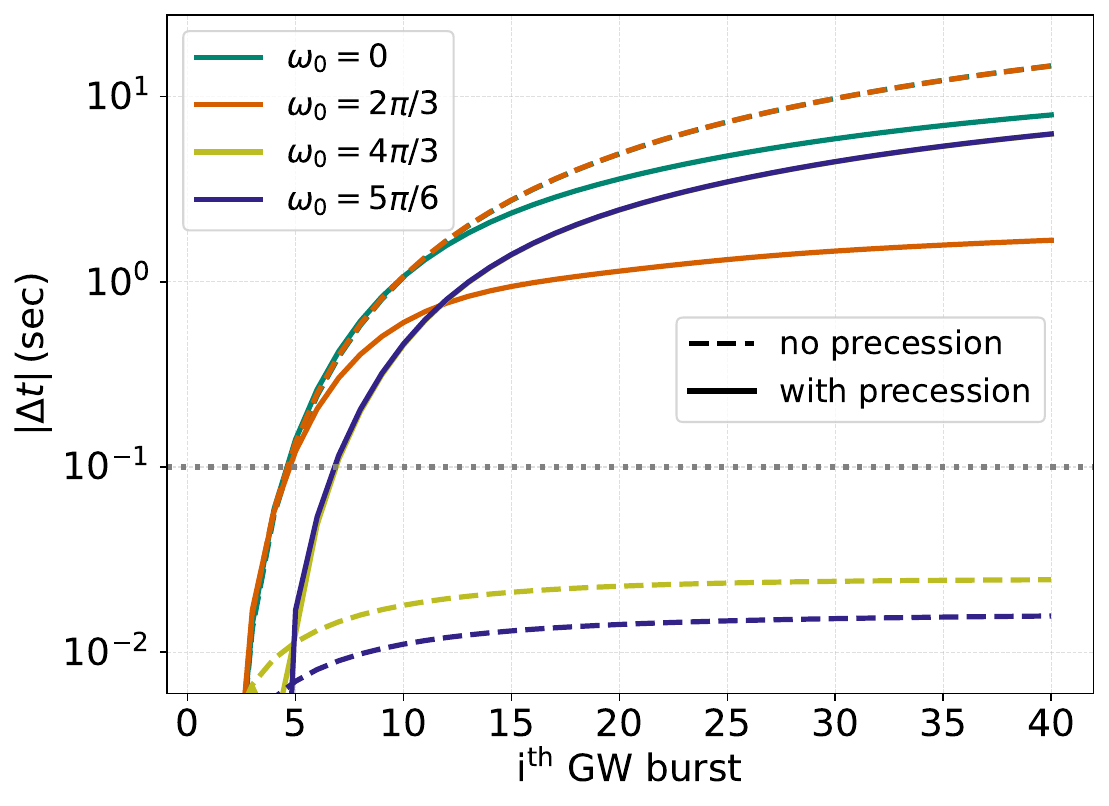}
    \caption{Absolute difference between the timings of the perturbed and unperturbed binary $|\Delta t|$ as a function of the number of GW bursts. Solid lines show the value of $\Delta t$ for a binary with 1PN precession, and dashed lines represents the same binary without 1PN precession. The horizontal dotted line indicates the detection threshold. The timings of the first burst are matched so that $\Delta t$ increases with time. We choose an equal mass binary with $m=30 M_{\odot}, e_0=0.99, p_0=100m$ and a tertiary of mass $m_3=10^6 m$ at distance $R=50 m_3$ and $V_{3,0}=\pi/3$. In the absence of 1PN precession, $\Delta t$ values show significant variation with $\omega_0$ and overestimate the phase shift between the perturbed and unperturbed binary. The yellow solid line is behind the blue line.} 
    \label{fig:w_vary}
\end{figure}   

\subsection{The role of 1PN precession on GW burst detectability}
Figure~\ref{fig:w_vary} shows the absolute difference in the timing $|\Delta t|$ of the same GW burst from the perturbed and unperturbed binary as a function of the number of GW bursts. The solid lines represent timing with 1PN precession, and the dashed lines indicate timing without precession. Different colors represent the different values of $\omega_0$. The horizontal dotted line shows $\Delta t = 0.1$ sec as the detection criterion. If $\Delta t > 0.1$ sec, the effect of the tertiary can be detectable. We choose an equal mass binary with $m=30M_{\odot}$, $e_0=0.99$, $p_0=100m$, $V_{3,0}=\pi/3$, $m_3 = 10^6 m$, and $R=50 m_3$. When precession is not included, the value of $\Delta t$ varies significantly and strongly depends on the initial value of $\omega_0$. For some values of $\omega_0$, the $\Delta t$ values fall significantly below the detection threshold. However, when the precession is taken into account, the values of $\Delta t$ are very close to each other for different values of $\omega_0$. In other words, when precession is taken into account, the dependence of burst timings is weaker on $\omega_0$ values. This is because the pericenter precession averages out any dependence on the initial phase of the binary. Note that for the cases when $\Delta t$ is above the detection threshold, unmodeled precession overestimates the value of $\Delta t$ as found in 3-body scattering simulations~\cite{Samsing:2024syt}.

\begin{figure}
    \centering
    \includegraphics[width=0.48\textwidth]{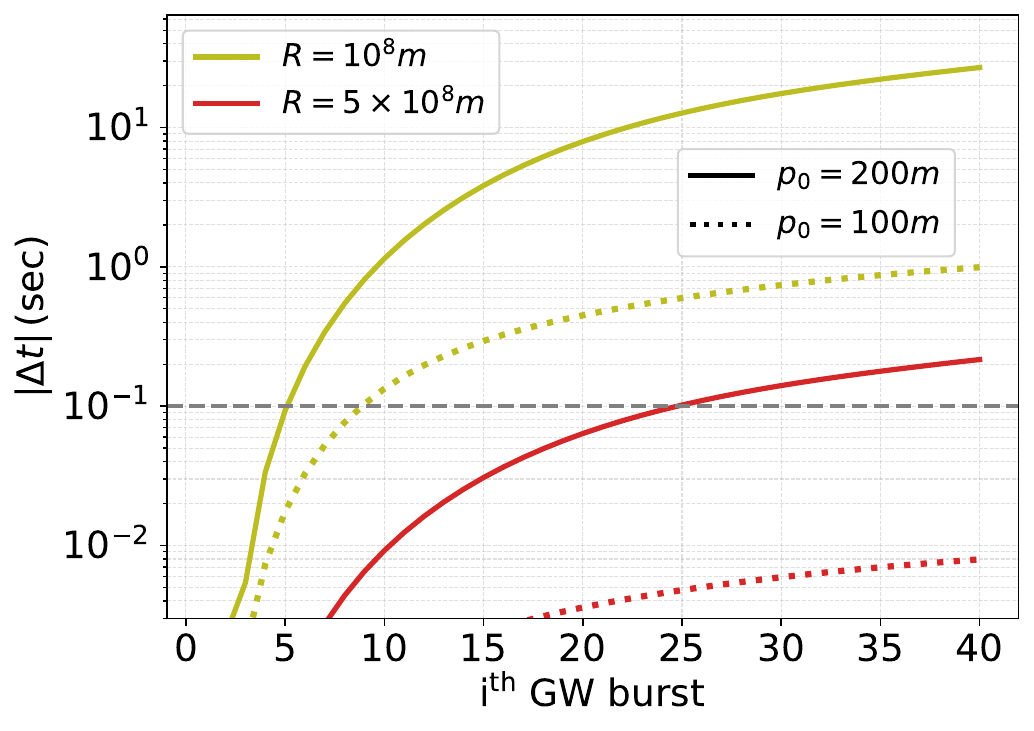}
    \caption{The $\Delta t$ as a function of the number of GW bursts. Different colors represent the different values of $R$. Solid lines indicate $\Delta t$ values for $p_0 = 200 m$, and dotted lines are for $p_0 = 100 m$. Other parameters are the same as those in Fig.~\ref{fig:w_vary}.}
   \label{fig:delta_t_R_p}
\end{figure}  

\subsection{Maximum separation of the third body to which the perturbation effects can be detected}\label{sec:maximum_distance}
Figure~\ref{fig:delta_t_R_p} illustrates $|\Delta t|$ as a function of the number of GW bursts for different values of $R$ and $p_0$. Each color corresponds to a different value of $R$. Solid lines represent results for an initial value of $p_0 = 200 m$, while dotted lines correspond to $p_0 = 100 m$. The value of $\Delta t$ increases as $p_0$ increases. This behavior arises because the perturbative contributions in Eqs.~\eqref{eq:p} \mbox{-} \eqref{eq:time} scale as $p^3$, and therefore for the fixed values of $m_3$ and $R$, a binary with a larger $p_0$ experiences a stronger perturbation. As a result, the third body's influence is more pronounced in wider binaries. For the same $m_3$ and $R$, a wider inner binary accumulates a larger phase shift over time due to the stronger cumulative influence of the perturbation. Quantitatively, for $p_0 = 200 m$, the timing effects induced by the third body can be detected out to distances of $R = 5 \times 10^8 m$. For $p_0 = 100 m$, the detectability range reduces to $R=10^8 m$. 

\subsection{The effect of the binary's center of mass motion on GW burst timing}\label{sec:COM_motion}
In addition to the tidal force from the third body, the binary's COM around the barycenter of the triple system changes the distance between the source and the detector, altering the arrival time of the bursts in the detector frame.  Generally Doppler-induced modulations are larger compared to tidal-induced modulations; however, Doppler-induced modulations can become suppressed depending on the location of the binary with respect to the detector. Under the assumption that the COM of the triple system is stationary, the observed timings of the GW bursts are given as~\cite{Isobel}
\begin{equation}\label{eq:observed_timing}
    t_n^{\rm obs} = t_n^{\rm int} - \bm{R} (t_n^{\rm int}) \cdot \bm{n}_{\rm obs} \,, 
\end{equation}
where $t_n^{\rm int}$ is the intrinsic timing of bursts and $\bm{R}(t^{\rm int}_i)$ is the separation vector between the inner binary and the third body. The vector $\bm{n}_{\rm obs}$ fixes the plane of the triple system with respect to the observer. It requires two angles: the inclination angle between the detector and the source plane $(\iota_3)$ and the angle of the line of nodes $(\Lambda_3)$.\footnote{Note that the angle $\Lambda_3$ is degenerate with the other two planar angles $V_3$ and $\omega_0$ and can be dropped from Eq.~\eqref{eq:observed_timing} with an appropriate choice of coordinate system. However, we keep $\Lambda_3$ for better visualization.} The vector $\bm{n}_{\rm obs}$ is given by 
\begin{equation}
    \bm{n}_{\rm obs} = [\sin \iota_3 \cos\Lambda_3, \sin\iota_3 \sin\Lambda_3, \cos\iota_3] \,.
\end{equation}
The observed time of the GW burst is 
\begin{equation}
    t_n^{\rm obs} = t_n^{\rm int} - R \sin\iota_3\cos[V_3(t_n^{\rm int}) + \Lambda_3] \,.
\end{equation}
Since we assume that the outer orbit does not evolve significantly over the observation time, the phase of the outer orbit can be approximated by $V_3 = \Omega_3 t_n^{\rm int}$, where $\Omega_3 = \left[(m+m_3)/R^3\right]^{1/2}$.

\begin{figure}
    \centering
    \includegraphics[width=0.48\textwidth]{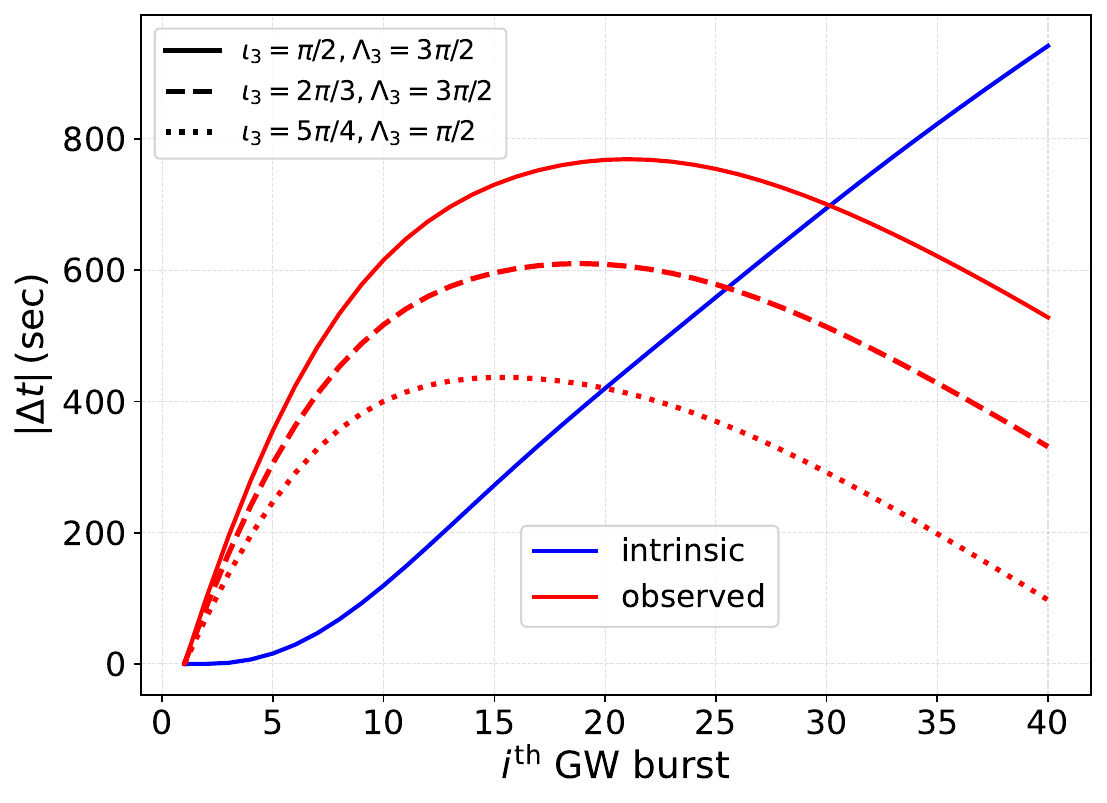}
    \caption{The $\Delta t$ for intrinsic GW timing (blue) and observed GW timing (red). We choose the same binary parameters as in Fig.~\ref{fig:w_vary} with $m_3=10^6 m$ and $R=10 m_3$. The observed timing can become suppressed depending on the location of the binary.}
   \label{fig:tidal_roemer}
\end{figure}    

To study the effects of Roemer delay on the burst timing, consider a system similar to a stellar mass BBH orbiting a SMBH. We consider the same binary as in Fig.~\ref{fig:w_vary} with $m_3=10^6 m$ and $R = 10 m$. Figure~\ref{fig:tidal_roemer} illustrates $\Delta t$ corresponding to the intrinsic GW timing and observed timing. While the intrinsic GW timing is independent of the location of the observer, the observed timing depends on the inclination angle $(\iota_3)$ and the angle of line of nodes $\Lambda_3$. Note that the intrinsic $\Delta t$ (purely due to tidal effects) increases as a function of time. The observed $\Delta t$ (with Doppler effects) can become suppressed depending on the location of the binary. Therefore, depending on the location of the binary, the tidal effects can be larger than the Doppler effects.  

\section{Summary and Conclusions}\label{sec:conclusion}
In this paper, we have studied the impact of conservative PN corrections on the dynamics of hierarchical triple systems consisting of a highly eccentric inner binary perturbed by a third body in a circular outer orbit. Highly eccentric binaries emit GW radiation in the form of GW bursts during each pericenter passage. The presence of a third body alters the timing of these GW bursts. We have incorporated 1PN corrections to the binary's orbital evolution and the timing of GW bursts. The timing model developed in this study presents an accurate analytical description of the orbital evolution of such triple systems. Our results show that 1PN-induced precession significantly alters the binary's orbital evolution.  

In particular, our study demonstrates that the inclusion of 1PN corrections significantly modifies the timing of GW bursts in both perturbed and unperturbed binaries. The timing deviations due to 1PN corrections become very large compared to the typical measurement uncertainty on burst timing ($\sim 0.1$ sec for Einstein Telescope). Moreover, the inclusion of 1PN corrections leads to the rapid change of the longitude of the pericenter, that induces distinctive modulation features in the semilatus rectum and eccentricity of the inner binary, which arise only when the system is both perturbed and precessing. Such signatures represent potential indicators of a nearby third body.

We further showed that neglecting 1PN corrections in the timing model leads to significant systematic bias in the mass and distance of the third body. The errors in the quantity $m_3/R^3$ become as large as $\mathcal{O}(10^{-1} \mbox{--} 10^6)$, introducing substantial bias in the estimation of the tertiary's properties. Additionally, in the absence of 1PN precession, the timing difference $\Delta t$ between perturbed and unperturbed binaries depends strongly on the initial argument of the pericenter, leading to inaccurate predictions for certain values of $\omega_0$. However, when 1PN precession is included, this dependence becomes significantly weaker. This has important implications for detection: the GW timing signature induced by a third body becomes more stable and robust against variations in the initial orbital configuration.  
 
All the above findings highlight the critical role of 1PN corrections in accurately modeling the timing and orbital evolution of eccentric binaries in hierarchical triple systems. Our results indicate that in the waveform modeling of hierarchical triple systems, 1PN corrections can lead to the potential breaking of degeneracies between model parameters -- leading to better parameter estimation.

While our model captures key relativistic effects in system dynamics, we have made a few simplified assumptions. We have neglected the dissipative effects beyond 2.5PN order, which may introduce additional, potentially non-negligible contributions to the binary's long-term evolution. We have also assumed that the outer orbit is circular and is coplanar with the inner binary, and we have omitted higher order PN corrections to the dynamics of the outer orbit. In a future study, it would be interesting to relax these assumptions and incorporate more realistic configurations -- including eccentric or inclined outer orbits, and higher order PN corrections -- to study their impact on system dynamics. Such extensions could reveal additional observational signatures~\cite{Hendriks:2024zbu, Hendriks:2024gpp}.

\section*{Acknowledgements}
This work was supported by the ERC Starting Grant No. 121817–BlackHoleMergs led by Johan Samsing, and by the Villum Fonden Grant No. 29466. J.T. acknowledges support from the Horizon Europe research and innovation programs under the Marie Sk\l{}odowska-Curie Grant Agreement No. 101203883. The Center of Gravity is a Center of Excellence funded by the Danish National Research Foundation under Grant No. 184.

\newpage

\appendix
\begin{widetext}

\section{2PN CORRECTIONS TO THE TIMING MODEL}\label{app:2PN corrections}
Since the leading order effects of pericenter precession appear at 1PN order, the contribution of next-to-leading (2PN) order precession in the timing model is subdominant. Nevertheless, we provide the 2PN corrections to the timing model for completeness. The 2PN acceleration is given by~\cite{PhysRevD.42.1123}
\begin{subequations}
\begin{align}    
    \bm{a}_{\rm 2PN} & = \frac{m}{r^2} \left(A_2 \bm{n} + B_2 \bm{v} \right)\,,
    \end{align}
where $A_2$ and $B_2$ are     
    \begin{align}
        A_2 &= - \frac{3}{4} (12+29 \eta) \left(\frac{m}{r} \right)^2 -\eta \left(3-4\eta \right) v^4 
         -\frac{15}{8}\eta \left(1-3\eta\right) \Dot{r}^4   \nonumber \\ 
         & +\frac{3}{2} \eta \left(3-4\eta\right) v^2 \Dot{r}^2  + \frac{1}{2} \eta \left(13-4\eta \right) \frac{m}{r} v^2 + \left(2+25 \eta + 2 \eta^2 \right) \frac{m}{r} \Dot{r}^2 \,, \\
        B_2 &=  \frac{\Dot{r}}{2} \Bigg[\eta \left(15+4\eta \right)v^2 - \left(4+41 \eta + 8 \eta^2 \right) \frac{m}{r} 
         - 3 \eta (3+ 2 \eta) \Dot{r}^2\Bigg] \,.
    \end{align}
\end{subequations}
The 2PN acceleration acts as a perturbing force, and we calculate its components $\mathcal{R}, \mathcal{S}$ and $\mathcal{W}$ in $\bm{n}$, $\bm{\lambda}$ and $\bm{e_z}$ basis, respectively, using Eq.~\eqref{eq:components_of_perturbing_force}, which are given as follows:
\begin{subequations}
\begin{align}
     \mathcal{R^{\rm 2PN}}  &= \frac{m^3}{64} \frac{(1+e\cos V)^2}{p^4} \bigg\{16 \left[-36 + \eta (-73 + 8 \eta) \right] +8 e \big[-144 -288 \eta + 80 \eta^2 + \eta e^2 (13+92 \eta) \big]\cos V \nonumber \\
     &+ 8 e^2 \big[-36 + \eta (-13 + 72 \eta) \big] + e^4 \eta (39 + 191 \eta) + e^2 \big[-288 -1192 \eta + 576 \eta^2 + 4 e^2 (-45 + 11 \eta) \big]\cos 2V \nonumber \\ 
     &  + e \eta (-456 + 160 \eta) \cos 3V + 3 e (-17 + 7 \eta) \cos 4V \bigg\} \,, \\ 
       \mathcal{S^{\rm 2PN}} &= \frac{m^3}{4} \frac{e (1+ e \cos V)^3 \sin V}{p^4} \bigg\{-8 + \eta \big[-52 -8 \eta + e^2 (21+ 2 \eta) \big] -2 e (4+11 \eta) \cos V + 3 e^2 \eta (3+2\eta) \cos 2V \bigg\}\,, \\
      \mathcal{W^{\rm 2PN}} & = 0 \,.
\end{align}
\end{subequations}
The secular change of orbital elements $p, e, w,t$ is obtained by integrating Eqs.~\eqref{eq:osculating_equations} over a complete orbit. The 2PN corrections to $\omega$ and $t$ are given by the following equations:
\begin{align}
    \omega_i^{\rm 2PN}  &= -\frac{\pi}{4} \left(\frac{m}{p_{i-1}} \right)^2 \bigg[8 \left(-7-5\eta+7\eta^2\right) 
     + e_{i-1}^2 \left(2-21\eta + 48 \eta^2 \right)\bigg] \,, \\
    t_i^{\rm 2PN} &= \nonumber  - \left(\frac{m}{p_{i-1}}\right)^2 \times \left(\frac{1-e_{i-1}^2}{8 e_{i-1}^2}\right) \bigg\{4 \left(1-\sqrt{1-e_{i-1}^2}\right) 
    \left(-4 + 14 \eta +\eta^2 \right) +  e_{i-1}^2 \bigg[8 \left(2+ \sqrt{1-e_{i-1}^2} \right) \\
    & +\eta \bigg(-53+131\sqrt{1-e_{i-1}^2}  -33 \eta + 71 \eta \sqrt{1-e_{i-1}^2} \bigg)  \bigg]  + e_{i-1}^4 \eta \left(-3+ 29 \eta \right) \bigg\} \,.
\end{align}
The above equations can be added to Eqs.~\eqref{eq:omega} and \eqref{eq:time} to obtain a timing model with 2PN corrections.
\begin{figure}[h]
    \centering
    \includegraphics[width=0.55\linewidth]{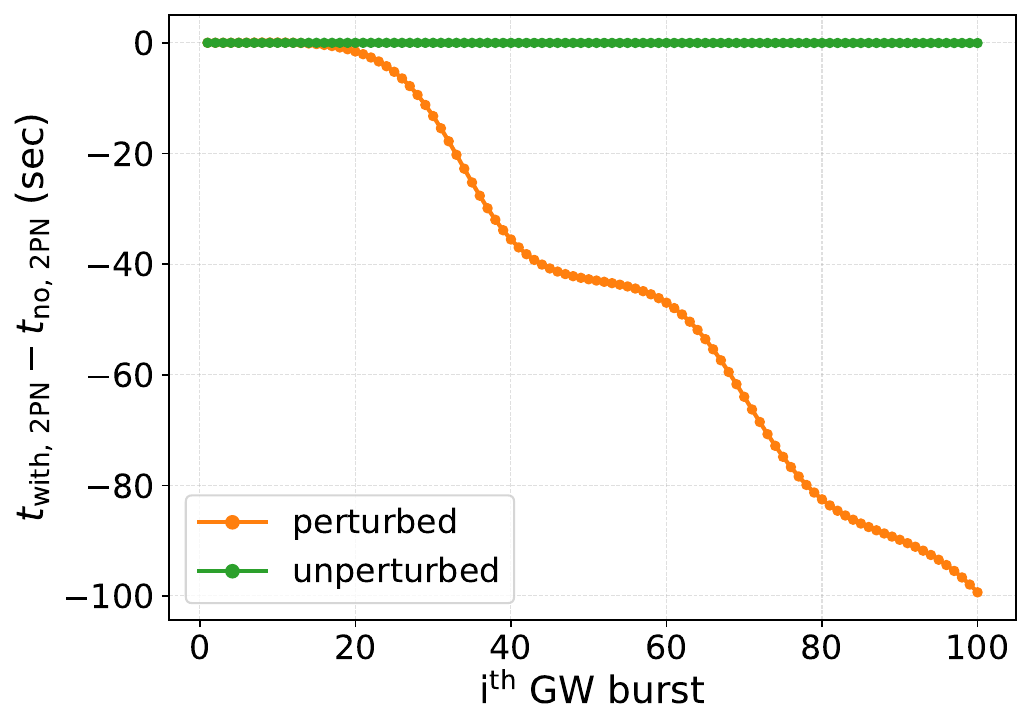}
    \caption{Difference in timings due to 2PN corrections as a function of the number of bursts. The system parameters are the same as those in Fig.~\ref{fig:timing}. For the perturbed binary, the difference between timings can reach up to $100$ sec at the $100^{\rm th}$ burst. }
    \label{fig:timing_2PN}
\end{figure}

In Fig.~\ref{fig:timing_2PN}, we plot the timing difference due to 2PN corrections. For the unperturbed binary, the timing difference is very small. For the perturbed binary, 2PN corrections make a large difference. The timing difference becomes as large as $\sim 100$ sec after the $100^{\rm th}$ burst. 

\end{widetext}

\bibliography{ref-list}
\end{document}